\documentclass[aps,prb,superscriptaddress,showpacs,twocolumn]{revtex4-1}
\usepackage{graphicx}
\usepackage{color}
\usepackage{txfonts}

\begin{document}

\title{Calculation of the graphene C 1\textit{s} core level binding energy}

\author{Toma Susi}
\email{toma.susi@iki.fi}
\affiliation{University of Vienna, Faculty of Physics, Boltzmanngasse 5, A-1090 Vienna, Austria}

\author{Duncan J. Mowbray}
\affiliation{Nano-Bio Spectroscopy Group and ETSF Scientific Development Centre, Departamento de F\'{i}sica de Materiales, Universidad del Pa\'{i}s Vasco UPV/EHU, E-20018 San Sebasti\'{a}n, Spain}
\affiliation{Donostia International Physics Center, Paseo Manuel de Lardizabal, 4. E-20018 Donostia-San Sebasti\'{a}n, Spain}
\author{Mathias P. Ljungberg}
\affiliation{Donostia International Physics Center, Paseo Manuel de Lardizabal, 4. E-20018 Donostia-San Sebasti\'{a}n, Spain}
\affiliation{Deparment of Physics, Phillips-University Marburg, Renthof 5, 35032 Marburg, Germany}

\author{Paola Ayala}
\affiliation{University of Vienna, Faculty of Physics, Boltzmanngasse 5, A-1090 Vienna, Austria}

\date{\today}

\begin{abstract}
X-ray photoelectron spectroscopy (XPS) combined with first principles modeling is a powerful tool for determining the chemical composition and electronic structure of novel materials. Of these, graphene is an especially important model system for understanding the properties of other carbon nanomaterials. Here, we calculate the carbon 1\textit{s} core level binding energy of pristine graphene using two methods based on density functional theory total energy differences: a calculation with an explicit core-hole ($\Delta$KS), and a novel all-electron extension of the delta self-consistent field ($\Delta$SCF) method. We study systematically their convergence and computational workload, and the dependence of the energies on the chosen exchange-correlation functional. The $\mathrm{\Delta}$SCF method is computationally more expensive, but gives consistently higher C 1\textit{s} binding energies. Although there is a significant functional dependence, the binding energy calculated using the PBE functional is found to be remarkably close to what has been measured for graphite.
\end{abstract}


\maketitle

X-ray photoelectron spectroscopy (XPS) is a powerful tool for studying the surface composition of materials. More recently, it has emerged as a particularly useful probe for low-dimensional carbon-based nanomaterials such as carbon fibers\cite{Takahagi88C}, thin films\cite{Merel1998}, nanotubes\cite{Larciprete05ASS,Ayala09PRB}, and graphene\cite{Hibino09PRB,Lizzit10NP}. Measured binding energies are often compared to molecular reference values to identify the corresponding atomic structures. However, for novel nanomaterials, appropriate references are often either not available, or it is unclear if they are directly applicable. Together with increases in computational power and method development, first principles modelling has gained more applicability for directly calculating the binding energies --- or at least the chemical shifts --- of desired atomic configurations \cite{Citrin78PRL,Egelhoff87SSR,Faulkner98PRL,Cole02JESRPa,Barinov09TJoPCC, Schiros12NL,El-Sayed13AN, Ruiz-Soria14AN,Cabellos12JPCC}.

The photoemission process can be conceptually divided into three basic steps. First, an X-ray photon is absorbed and transfers its energy to a single core electron, creating a photoelectron. Then, this electron makes its way to the surface of the material. Finally, the electron escapes from the surface into the vacuum. Experimentally, the need for knowing the work function of the material in the last step is bypassed by referencing the binding energies to the Fermi level of the material, which is a well-defined procedure for systems without a band gap. 

For calculating core level binding energies, two types of methodologies are typically applied: the so-called initial state and final state methods \cite{Cabellos12JPCC}. In the initial state methods, only the energy level of the core electron before ionization is considered, often by simply calculating its Kohn--Sham (KS) orbital eigenvalue using density functional theory (DFT), referenced to the Fermi level. This is typically accomplished by explicitly including the core level via an all-electron (ae) calculation. Initial state methods have the advantage that the KS eigenenergies may be calculated for all atoms of the system within a single calculation. The justification for this procedure is a linearization around the ground state of Janak's theorem \cite{Janak78PRB}, which states that the orbital energy is the derivative of the total energy with respect to the orbital occupation. However, the absolute values of carbon core levels are typically underestimated by about 10\% by DFT \cite{Cabellos12JPCC}, partly because core-hole relaxation is disregarded within this approximation \footnote{This approximation also neglects the derivative discontinuity of the exchange-correlation functional at whole occupation numbers\cite{Perdew83PRL,Sham83PRL}}.

In the final state methods, the core-hole is explicitly included in a second calculation, and the electronic structure relaxed in its presence. The binding energy of the core electron is then computed from the total energy difference between the excited state with the core-hole ($E_{ex}$) and the initial ground state configuration ($E_{gs}$). Since only total energy differences are used in the calculation, final state methods take advantage of DFT's high level of accuracy with respect to total energies, and avoid the well-known problems of describing energy levels using KS eigenvalues. However, a separate calculation must be performed for each atom of interest. The Slater transition state method should also be mentioned, where the excitation energy is calculated from the orbital energy differences in a state halfway between the initial and final states, that is, with a non-physical half-core-hole. However, this method is in general not as accurate as the final state methods, and it shares their complication of requiring an explicit core-hole.

Modeling the final state with a core-hole is significantly more challenging than a ground state calculation. If the core-hole is introduced via a projector augmented-wave (PAW) dataset (i.e., a PAW setup) or within an atomic pseudopotential, the atom becomes charged in the final state. A periodic ``bare core-hole'' calculation would require a huge supercell to properly include this charge distribution \cite{Despoja12PRB}. Further, for low-dimensional materials, the long-ranged Coulomb interaction introduces an additional slow convergence of the total energy with the amount of vacuum \cite{Komsa14PRX}. These issues may be partly addressed by explicitly including an extra electron charge within the conduction band of the material, resulting in a so-called ``screened core-hole''. However, using a PAW dataset or a pseudopotential does not allow the other core electron(s) to relax, which may limit the accuracy of the absolute binding energies \cite{Susi14BJN}. Although a rigid shift can be applied to align the calculated values with experiment, this assumes that the effect of core-hole relaxation is of identical magnitude for every atom of interest --- which can be \textit{a priori} uncertain for atoms of different elements. Thus, accurate absolute values from a physically motivated calculation are of great practical interest.

As the prototypical low-dimensional carbon nanomaterial, graphene \cite{Novoselov04S,Novoselov05N,Geim07NM,CastroNeto09RMP} is useful for understanding the structure and often also the properties of other interesting materials, such as carbon nanotubes. Significant efforts have been directed towards modifying its properties, such as opening a band gap or tuning the carrier concentration, by chemical functionalization \cite{Elias09ISTPI,Leconte10AN,Hossain12NC} or by heteroatom doping \cite{Wei09NL,Terrones12RPP}. For such studies, a chemically sensitive quantitative probe like XPS is a vital tool for discerning the amount and bonding of dopant atoms or functional groups.

Here, we calculate the C 1\textit{s} core level binding energy of pristine graphene using two methods based on DFT: a delta Kohn--Sham ($\mathrm{\Delta}$KS) calculation using a PAW-dataset including an explicit core-hole, and a novel application of the delta self-consistent field ($\mathrm{\Delta}$SCF) method including the core levels within an all-electron calculation (see Ref. \footnote{On the naming of the methods: $\mathrm{\Delta}$Kohn--Sham and $\mathrm{\Delta}$SCF are used synonymously in the literature. SCF was used to denote the Hartree--Fock method in the quantum chemistry community, which is why the original DFT total energy difference method dealing with core states was dubbed $\mathrm{\Delta}$KS.} for a note on the nomenclature). We study the convergence and computational workload of both methods, the functional dependence of the energies, and show how the magnetic moment affects the $\mathrm{\Delta}$KS results.


Our DFT calculations were performed with the grid-based projector augmented-wave simulation package \textsc{gpaw} \cite{Mortensen05PRB,Enkovaara2010}. Exchange and correlation were estimated by the Perdew-Burke-Ernzerhof (PBE) generalized gradient approximation \cite{Perdew96PRL}, and the LDA\cite{Perdew92PRB}, PW91\cite{Perdew92PRB}, revPBE\cite{Zhang98PRL} and RPBE\cite{Hammer99PRB} functionals tested in selected cases. We applied periodic boundary conditions in orthorhombic unit cells of 2 to 11 elementary lattice units, yielding supercells with 8 to 242 carbon atoms. Monkhorst-Pack \cite{Monkhorst76PRB} 3$\times$3$\times$1, 5$\times$5$\times$1 or 7$\times$7$\times$1 \textit{k}-point meshes were applied depending on the cell size (yielding 3, 5 and 8 \textit{k}-points in the irreducible part of the Brillouin zone). The relaxed graphene lattice parameter was $a =$~2.443 \AA, while the grid spacing sufficient for convergence of the C 1\textit{s} was $h\approx0.19$ $\AA$ (spacings down to 0.10 $\AA$ were tested).

In the $\mathrm{\Delta}$KS total energy differences method \cite{Susi14BJN,Ljungberg11JESRP}, the core level binding energy is the total energy difference between a first core ionized state and the ground state in a spin-polarized calculation. To make the unit cell charge-neutral, a compensating electron charge is introduced into the conduction band. This is a good approximation for metals (including graphene) where core-hole screening is efficient. We additionally investigated the effect of different magnetic moments of the final state. For a singlet,
we initialized the magnetic moment of the core-hole atom to 1.0 Bohr magnetons (counting valence electrons, with the core-hole in spin up) and fixed the total magnetic moment, and also ran fixed calculations with -1.0 Bohr magnetons (triplet). Otherwise the magnetic moment was allowed to relax freely.

\begin{figure}[]%
\center{\includegraphics[width=0.35\textwidth]{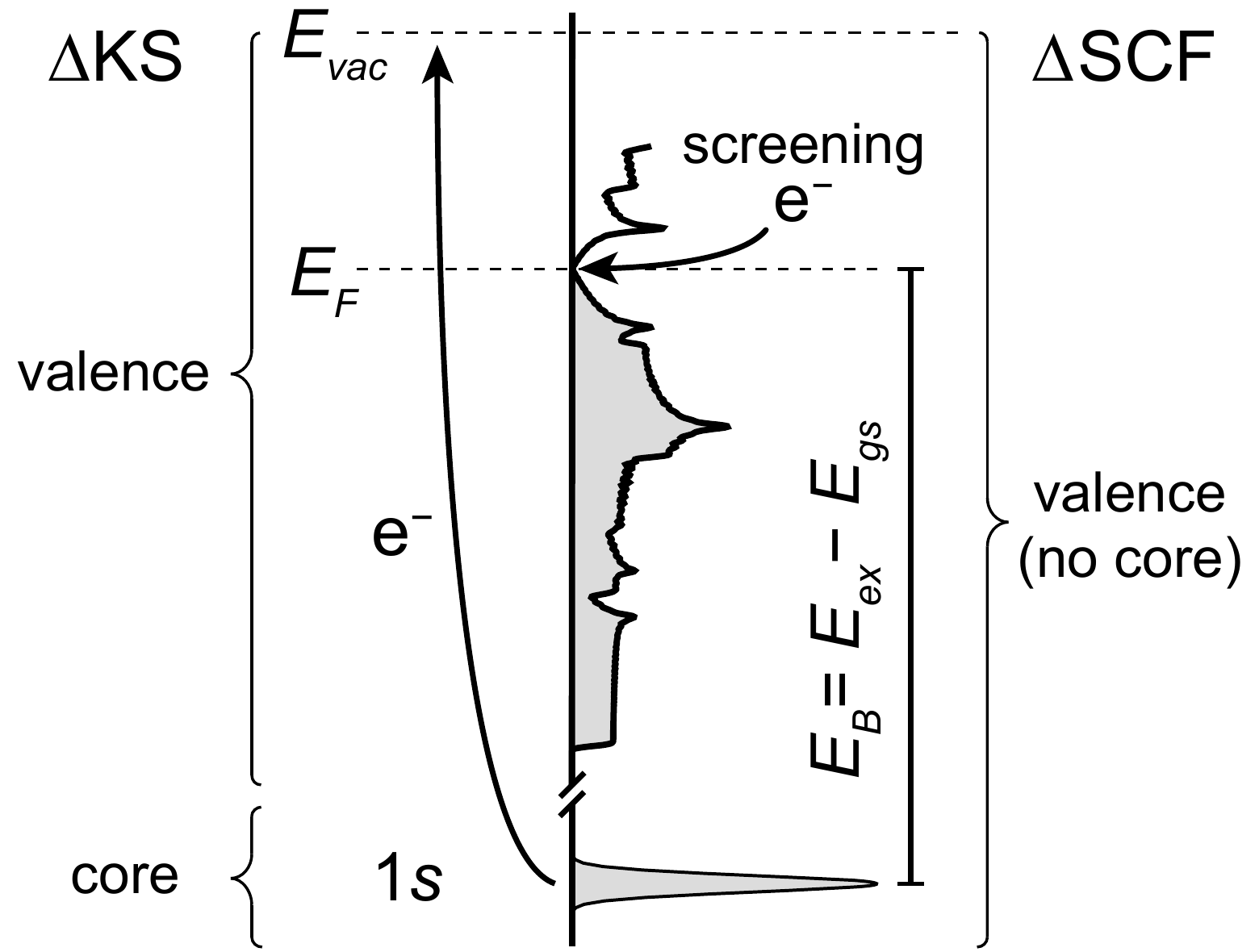}}
\caption{A schematic illustration of the core level binding energy ($E_B$) of graphene, which we calculate as the difference between the excited state ($E_{ex}$) and ground state ($E_{gs}$) total energies. In both the $\mathrm{\Delta}$KS and the $\mathrm{\Delta}$SCF excited states, one electron ($e^-$) is removed from the 1\textit{s} core state to vacuum ($E_{vac}$), and a compensating electron charge is introduced at the Fermi level ($E_F$). However, in $\mathrm{\Delta}$KS, the core is described by a PAW dataset including an explicit core hole, while in $\mathrm{\Delta}$SCF, all electrons are included in the valence and the core hole described by subtracting the density of a spin-up carbon 1\textit{s} orbital.}
\label{schematic}
\end{figure}

We then turned to the delta self-consistent field ($\mathrm{\Delta}$SCF) method implemented\cite{Gavnholt08PRB} in \textsc{gpaw}. As a modification to include core levels in the calculation, we used so-called ``pseudoatom'' all-electron datasets. In this recently implemented feature, the core states are included in the valence, enabling an explicit ae calculation within the PAW scheme \cite{Ojanpera14PRB} (note that this is different from the relaxed core method of Marsman and Kresse \cite{Marsman06}). In a $\mathrm{\Delta}$SCF calculation, the density of a specified orbital $\varphi_a(r)$ (in this case a spin-up carbon 1\textit{s} orbital) is subtracted from the total density in each step of the self-consistency cycle. As in the $\mathrm{\Delta}$KS method, the missing core charge is compensated by an extra electron in the conduction band. Figure \ref{schematic} illustrates the methods schematically. Finally, we tested the influence of using ae datasets on other atoms in the system in both the $\mathrm{\Delta}$KS and the $\mathrm{\Delta}$SCF calculations.

Turning now to our results, we first studied the influence of the compensating charge in the $\mathrm{\Delta}$KS method by calculating the C 1\textit{s} energy of a charged 9$\times$9 graphene supercell as a function of the perpendicular separation of the periodic images of the graphene plane (along the $z$-axis in our geometry). We found convergence to be very slow, not reaching a constant value even for a separation of 50 \AA. Furthermore, the calculations trended towards a significantly too high binding energy (288.15 eV). However, when the system was made charge-neutral, only 8 \AA~of vacuum was enough to converge the C 1\textit{s} energies. (For the charge-neutral unit cell, non-periodic boundary conditions in the $z$-direction yielded no difference to a periodic calculation.)

Concluding thus that the extra charge is needed, we considered the convergence of the energies as a function of the supercell size and the number of \textit{k}-points in the calculation. For even the smallest 2$\times$2 supercell, a \textit{k}-point mesh of 7$\times$7$\times$1 (mesh density $\Delta k <$~0.2~\AA$^{-1}$) was enough to converge both the ground and excited state energies to within 1 meV per atom. However, although the absolute changes in energy were not large, convergence of the excited state energy was found to be rather slow as a function of system size. This is likely due to the long-range Coulomb interaction between periodic images of the core-hole, which destabilizes the final state and artificially increases the excited state energies. Overall, for the largest unit cells (9$\times$9 and above), we found a $k$-point mesh of 3$\times$3$\times$1 to be sufficient for full convergence.

\begin{figure}[]%
\center{\includegraphics[width=0.48\textwidth]{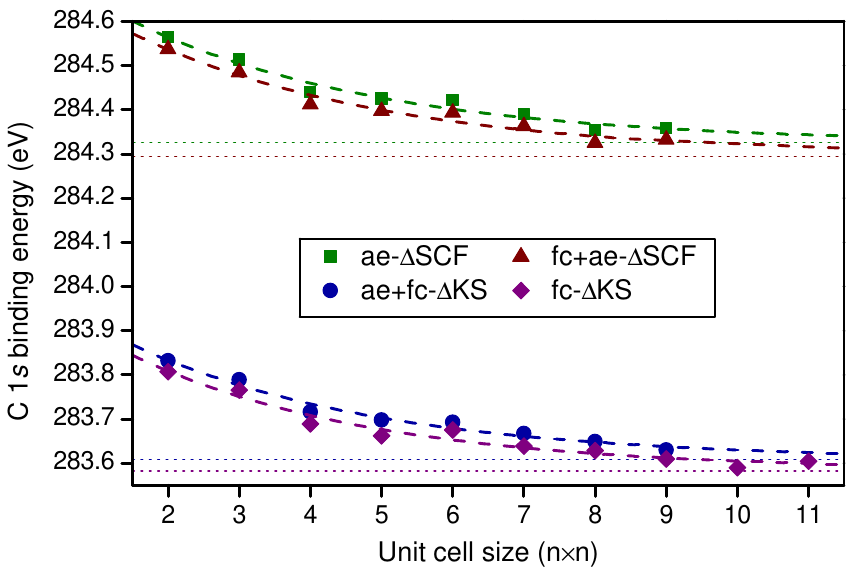}}
\caption{(color online.) The graphene C 1\textit{s} binding energy as a function of supercell size calculated with the $\mathrm{\Delta}$SCF and $\mathrm{\Delta}$KS methods using frozen core (fc) and/or all-electron (ae) PAW datasets as described in the text. A sufficient number of \textit{k}-points were employed throughout. Decaying exponential fits yield asymptotic limits (dotted horizontal lines) representing extrapolations for fully converged values (Table \ref{table1}).}
\label{C1s}
\end{figure}

In Fig. \ref{C1s}, we have plotted the \textit{k}-point converged binding energies for each unit cell size from the $\mathrm{\Delta}$KS and the $\mathrm{\Delta}$SCF calculations. For each method, we have fitted the data with decaying exponentials, whose $y$-offsets give estimates for fully converged C 1\textit{s} energies. We see that for the largest computationally tractable 11$\times$11 unit cell containing 242 carbon atoms, the $\mathrm{\Delta}$KS values are converged to within 50 meV. For more standard sizes like the 6$\times$6 cell, both methods are about 100 meV higher than the fully converged value. When we included ae datasets on all other atoms in the core-hole calculation (ae+fc-$\mathrm{\Delta}$KS), the converged value was raised by only 30 meV compared to the all-fc calculation. Conversely, when we performed $\mathrm{\Delta}$SCF calculations with an ae dataset just on the target atom and normal fc datasets on other atoms (fc+ae-$\mathrm{\Delta}$SCF), we see that the ae-$\mathrm{\Delta}$SCF values are systematically only 30 meV higher in energy. Thus the relaxation of core electrons on neighboring atoms does not appear to be significant. Furthermore, fixing the magnetic moment to the singlet value in the $\mathrm{\Delta}$KS calculation was found to raise the converged value by about 0.1 eV, with the triplet being about 30 meV lower than this.

\begin{table}
  \caption{Converged graphene C 1\textit{s} binding energies calculated with the methods described in the text using the PBE functional. The last two columns give the CPU time scaling $\alpha kN^{\beta}$ prefactors and exponents. \label{table1}}
\begin{ruledtabular}
   \begin{tabular}{lclccc}
      & C 1\textit{s} & \multicolumn{2}{c}{Scaling}\\
Method&(eV)&$10^{-5}\alpha$ & $\beta$\\\hline
     fc-$\mathrm{\Delta}$KS & 283.58 & 0.06& 2.86\\
     ae+fc-$\mathrm{\Delta}$KS & 283.61 & 36& 2.14\\
     fc+ae-$\mathrm{\Delta}$SCF & 284.29 & 14&  drr2.31\\
     ae-$\mathrm{\Delta}$SCF & 284.33 & 7.2 & 2.59\\
\end{tabular}
\end{ruledtabular}
\end{table}

Experimentally, the reference value of the C 1\textit{s} binding energy of graphite is 284.42 eV \cite{Prince00PRB, Speranza07JAP}. For graphene, values found in the literature range from 283.97 eV for graphene on Pt(111) \cite{Rajasekaran13PRL, Preobrajenski08PRB}, 284.15 on Ir(111) \cite{Preobrajenski08PRB, Lizzit10NP}, 284.2 eV on Au-intercalated Ni(111) \cite{Haberer10NL,Haberer11AM}, 284.47 eV for suspended few-layer graphene \cite{Scardamaglia14C}, 284.6 eV on hydrogen-intercalated SiC \cite{Riedl09PRL}, 284.7 eV on Ni(111) \cite{Grueneis09NJP}, to 284.8 eV on SiC \cite{Emtsev08PRB, Hibino09PRB}. While it is thus clear that charge transfer from and screening by the substrate affect the measurements significantly, the exact value for freestanding single-layer graphene has not been fully established.
 
Taking the graphite value as the experimental reference against which to evaluate the data in Fig. \ref{C1s}, we can see that the PBE $\mathrm{\Delta}$KS underestimates the binding energy by about 0.8 eV, as we observed before\cite{Susi14BJN}. However, when using the $\mathrm{\Delta}$SCF method, the relaxation of the other core electron of the target atom is included in the description, unlike with the frozen-core (fc) PAW datasets. With the fully ae $\mathrm{\Delta}$SCF method, we get a converged C 1\textit{s} energy of 284.33 eV, constituting only a 0.03$\%$ difference to the experimentally reported graphite binding energy. (Although currently only possible in the $\mathrm{\Delta}$KS method, fixing the spin state of the extra charge to a singlet would likely have a similar magnitude effect also for the $\mathrm{\Delta}$SCF value, raising the C 1\textit{s} energy by a further 0.1 eV.)

However, the near-perfect agreement with the graphite measurement that results should be considered fortuitous since the choice of the exchange-correlation functional was found to affect the energies by several tenths of an eV. To see this, we selected the fc+ae-$\mathrm{\Delta}$SCF and fc-$\mathrm{\Delta}$KS methods, and looked at the C 1\textit{s} values calculated for the 9$\times$9 unit cell (Table \ref{table2}). We see that while LDA gives drastically lower energies, the other functionals are within 0.7 eV of each other, with PBE notably being the next lowest in energy for both methods, consistent with results obtained for molecules \cite{Takahashi04JCP}. Thus, while the functional dependence can be used as an estimate for the uncertainty in our calculated values, the functional that reproduces the experimental value best may be considered the most useful for core level calculations using this methodology. We should also note that the total energy (including atomic reference energies) of the fc+ae-$\mathrm{\Delta}$SCF ground state was consistently about 0.25 eV lower and the excited state about 0.3 eV higher than the corresponding fc-$\mathrm{\Delta}$KS ones. Although calculations with a finer grid lowered both ground and excited state energies, this did not affect the total energy differences appreciably.

\begin{table}
  \caption{The functional dependence of our calculated C 1\textit{s} energies with the fc-$\mathrm{\Delta}$KS and fc+ae-$\mathrm{\Delta}$SCF methods in the 9$\times$9 supercell.\label{table2}}
\begin{ruledtabular}
   \begin{tabular}{lccc}
      & fc-$\mathrm{\Delta}$KS & fc+ae-$\mathrm{\Delta}$SCF\\
     XC & C 1\textit{s} (eV) & C 1\textit{s} (eV) & Difference (eV)\\\hline
     LDA & 280.90 & 281.32 & 0.42 \\
     PBE & 283.77 & 284.33 & 0.58 \\
     PW91 & 284.02 & 284.69 & 0.71 \\
     revPBE & 284.15 & 284.84 & 0.69 \\
     RPBE & 284.30 & 284.99 & 0.69 \\
\end{tabular}
\end{ruledtabular}
\end{table}

\begin{figure}[]%
\center{\includegraphics[width=0.48\textwidth]{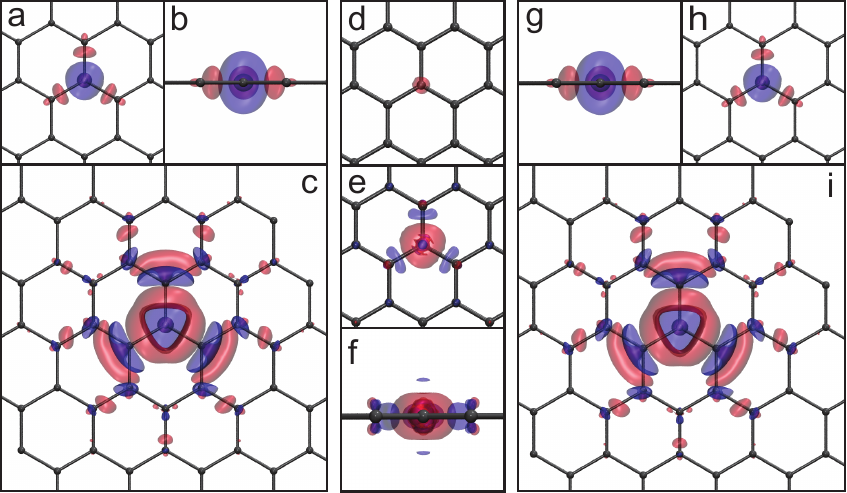}}
\caption{(color online.) All-electron charge density difference isosurfaces calculated in the 6$\times$6 supercell between the (a-c) $\mathrm{\Delta}$KS excited and ground state (side view in b), (g-i) $\mathrm{\Delta}$SCF excited and ground state (side view in g), and (d-f) $\mathrm{\Delta}$KS excited state and $\mathrm{\Delta}$SCF excited state (side view in f). Positive values are denoted in red and negative in blue (isovalues $\pm$0.1 (a,b,g,h), $\pm$0.01 (c,i), $\pm$0.02 (d), and $\pm$0.0015 e/$\AA^3$ (e,f)).}
\label{coreholes}
\end{figure}

We further reconstructed the all-electron densities for each calculation, and computed differences between the  $\mathrm{\Delta}$KS (Figure \ref{coreholes} a-c) and $\mathrm{\Delta}$SCF excited states and ground states (Figure \ref{coreholes} g-i), and between the two excited states (Figure \ref{coreholes} d-e). The isosurfaces displaying the differences between the excited state and ground state charge densities in each method look very similar, confirming that the forced occupation of the core orbital in the $\mathrm{\Delta}$SCF method reproduces the general features of the better tested frozen core-hole dataset. Only by looking at the difference between the two excited state densities plotted at low isovalues (Figure \ref{coreholes} d-f), subtle differences between the two methods can be seen near the core-hole atom.

Finally, we considered the computational effort required to complete each calculation (total running time multiplied by the number of cores). The computational time scales theoretically with the number of atoms \textit{N} in the supercell and with the number of \textit{k}-points in the irreducible part of the Brillouin zone. We can thus model the CPU time data as $\alpha kN^{\beta}$ and use the scaling prefactors $\alpha$ and exponents $\beta$ as given in Table \ref{table1} to compare the different methods. As an example of actual times, for an 8$\times$8 unit cell of 128 atoms, the calculations with the ae-$\mathrm{\Delta}$SCF, fc+ae-$\mathrm{\Delta}$SCF, ae+fc-$\mathrm{\Delta}$KS, and fc-$\mathrm{\Delta}$KS methods took 20.7, 10.0, 11.4 and 0.76 CPU-hours to complete, respectively. Thus, we can see that the fc-$\mathrm{\Delta}$KS calculations are much faster than the other methods.

To conclude, our results indicate that prohibitively large unit cells are required to completely converge the C 1\textit{s} core level binding energy of graphene using DFT calculations with periodic boundary conditions. However, for larger system sizes, convergence within 50 meV is reached and the underestimation is systematic. Thus, when choosing a size for the computational unit cell, one can balance considerations of computational efficiency (when a large number of systems or target atoms need to be simulated) with for example the requirement of having a realistic concentration of defects or dopants. However, although computationally cheap, the $\Delta$KS calculations underestimate the experimentally expected value by about 0.8 eV. By performing physically motivated $\Delta$SCF calculations using all-electron datasets, systematically higher binding energies were obtained, although the exact value was found to be sensitive to the chosen exchange-correlation functional. Nonetheless, the PBE functional gives a C 1\textit{s} binding energy that is remarkably close to the experimental value.

\begin{acknowledgments}
We acknowledge generous grants of computing time from the Vienna Scientific Cluster. T.S. was supported by the Austrian Science Fund (FWF) through grant M 1497-N19, by the Finnish Cultural Foundation, and by the Walter Ahlstr\"{o}m Foundation. D.J.M.\ acknowledges funding through the Spanish ``Juan de la Cierva'' program (JCI-2010-08156), Spanish Grants (FIS2010-21282-C02-01) and (PIB2010US-00652), and ``Grupos Consolidados UPV/EHU del Gobierno Vasco'' (IT-578-13). M.P.L. was supported by the German DFG Collaborative Research Centre (Sonderforschungsbereich) SFB 1083.
\end{acknowledgments}

\bibliography{xps_review}

\end{document}